\documentclass[structabstract]{aa}  

\usepackage{graphicx}
\usepackage{txfonts}
\usepackage{natbib}
\bibpunct{(}{)}{;}{a}{}{,} 

\begin{document}

 \title{Optical and infrared observations of the Crab Pulsar and its 
nearby knot\thanks{Based on observations made with ESO Telescopes 
under programme ID 072.D-0029.
Tables 3 and 4 and Figure 4 are only available in electronic form via 
http://www.edpsciences.org
 }}

   \author{A.~Sandberg
          \and
          J.~Sollerman
          }
\institute{The Oskar Klein Centre, Department of Astronomy, Stockholm University, AlbaNova, 106 91 Stockholm, Sweden.\\
              \email{jesper@astro.su.se}
             }

  \abstract
    {}
 {
We study the spectral energy distribution (SED) 
of the Crab Pulsar and its nearby knot 
in the optical and in the infrared (IR) regime. 
We want to investigate how the contribution from the knot 
affects the pulsar SED in that regime, and examine the evidence for
synchrotron self-absorption in the IR.
We also draw the attention to the predicted secular decrease 
in luminosity of the Crab Pulsar, and 
attempt to investigate this with CCD observations.
}
{
We present high-quality $UBVRIz$, 
as well as adaptive optics $JHK_sL'$ photometry, 
achieved under excellent conditions with the FORS1 and NAOS/CONICA 
instruments at the VLT.
We combine these data with re-analyzed archival Spitzer Space Telescope
data to construct a SED for the pulsar, 
and quantify the contamination from the knot. 
We have also gathered optical imaging data from 1988 to 2008 from several telescopes 
in order to examine the predicted secular decrease in luminosity.
}
 {
For the Crab Pulsar SED we find a spectral slope of 
$\alpha_\nu = 0.27 \pm 0.03$ in the optical/near-IR regime, 
when we exclude the contribution from the knot. 
For the knot itself, 
we find a much redder slope of $\alpha_\nu = -1.3 \pm 0.1$. 
Our best estimate of the average decrease in luminosity for the 
pulsar is $2.9 \pm 1.6$ mmag per year.}
{
We have demonstrated the importance of the nearby knot in precision 
measurements of the Crab Pulsar SED, in particular in the near-IR. 
We have scrutinized the evidence for the traditional view of a 
synchrotron self-absorption roll-over in the infrared, 
and find that these claims are unfounded.
We also find evidence for a secular decrease in the optical light 
for the Crab Pulsar, in agreement with current pulsar spin-down models.
However, although our measurements of the decrease significantly improve 
on previous investigations, the detection is still tentative.
We finally point to future observations that can improve the situation significantly.
}


\headnote{For publication in A\&A}
\titlerunning{The Crab Pulsar in optical and near-IR}
\authorrunning{Sandberg \& Sollerman}

\maketitle

%

\section{Introduction}

The Crab Nebula and its pulsar are among the most extensively studied objects 
in astronomical research. The nebula is the remnant of a supernova that was
observed in 1054 A.D. Even though well studied, a complete understanding of 
the supernova that exploded, and the nature of the nebula, is still missing
\cite[e.g.,][]{sollerman01,kitaura06,hester08,tziamtzis09}.

In this paper we focus on the pulsar itself and its immediate environment. 
The Crab Pulsar has been utilized for decades for testing
high-energy astrophysics. But the exact nature of pulsars, 
including their emission mechanism, is far from fully understood 
\cite[e.g.,][]{romani00}. 
Even in the well studied optical and near-infrared (IR) domain 
there remain open issues regarding e.g., 
the origin of the off-pulse polarisation 
\citep{shearer02,slowikowska09,mignani09}, 
and the exact spectral-energy-distribution (SED) of the non-thermal emission
\citep{sollerman01,oconnor05,temim06} including the 
influence of the nearby knot, located approximately $0\farcs6$ ($\sim$1000 AU) 
from the pulsar \citep{hester95,sollerman03}.

We present new optical and near-IR observations of the Crab Pulsar and 
its immediate environment in order to trace the SED into the infrared, 
and to investigate the
spectral shape of the emission from the knot. 
These observations were obtained with the ESO Very Large Telescope (VLT). 
We also present an attempt to determine the secular decrease of the 
optical emission by a study of optical CCD imaging 
data obtained over the past 20 years.

The paper is organized as follows. 
In Sect.~\ref{observations} we
describe the optical observations obtained under excellent conditions
with the FORS1 instrument on the VLT. We also describe the adaptive
optics NAOS/CONICA near-IR data from VLT that extend our observations all the
way into the $L'$ band.  
To extend the SED even further, we also include and 
re-analyze archival data from the Spitzer Space Telescope. In
Sect.~\ref{results} we present the SEDs we derive for the pulsar and
for the nearby knot, and these results are further discussed in
Sect.~\ref{discussion}.  
We also present a complementary investigation
of the optical emission from the Crab Pulsar in Sect.~\ref{secular},
where we discuss the evidence for a secular decrease of the $V$-band
emission from the pulsar using archival data from many telescopes.
The paper ends with some conclusions and an outlook in
Sect.~\ref{conclusions}.

\section{Observations}\label{observations}

\subsection{Optical photometry with FORS1}\label{optical}

The Crab Nebula was observed on December 18, 2003, with the FORS1 instrument 
mounted on UT1 at the VLT. The observations were conducted in service mode
to ensure excellent observing conditions. We used the
high-resolution mode of the camera, utilizing 0.1 arcsec per pixel. 
For each of the $UBVRIz$ pass-bands, we obtained three dithered exposures. 
No previous photometry has been reported for the pulsar in the $z$ band, 
and this was included in order to
tie together the optical and near-IR observations. 

The seeing, as measured from the final images, 
was $\sim$0$\farcs$6-0$\farcs$7 in the $UBVR$ exposures and
$\sim$0$\farcs$7-0$\farcs$8 for the $I$ and $z$ bands. 
For photometric calibrations, the white dwarf spectrophotometric standard star 
GD\,71 \citep{bohlin95} was observed in the $UBRIz$ 
bands\footnote{The $V$-band exposure of GD\,71 was unfortunately saturated.} 
immediately after the pulsar observations, and at a similar airmass. 
For the photometric calibration, we have also made use of 
observations of stars in the 
Rubin\,152 field \citep{landolt92} obtained during the same night.

All FORS1 data were reduced in a standard way, 
including bias subtraction and flat fielding.
We measured the Crab Pulsar and other stars in the field using 
aperture photometry with
an aperture of 8 pixels. The field stars
were used for calculating an aperture correction for the pulsar. 
The background 
was estimated with an annulus from 25 to 35 pixels in the Crab
field, using the centroid sky algorithm.

Point-spread Function (PSF) subtractions were performed on all FORS1 images, 
using the {\tt DAOPHOT} algorithm package as implemented in IRAF.
\footnote{IRAF is distributed by the National Optical Astronomy Observatory, which 
is operated by the Association of Universities for Research in Astronomy (AURA) 
under cooperative agreement with the National Science Foundation.} 
This was done in order to remove the 
pulsar emission and thereby reveal the nearby knot. 
To reduce the effect of residuals from the pulsar, 
the knot photometry was performed with an aperture of only 5 pixels, and an
aperture correction 
from 5 to 25 pixels was estimated from the same
field stars as for the pulsar.
Although the knot is an extended 
structure, the true width is only $\sim 0\farcs2$ (Sect.~\ref{discussion}),
and the knot can be approximated as a point source for this purpose.

Since no standard $z-$band magnitude exists for GD\,71, it was estimated by
integrating the spectrum of GD\,71 under the FORS1 filter profile using 
the IRAF task {\tt sbands}. 
For normalization, the same procedure was followed for 
the spectrum of Vega, and defining its magnitude to zero. 
When applying this procedure to GD\,71, 
we obtain a $z-$band magnitude of 
$z=13.42\pm0.03$ for this star. 
We note that using the same procedure to estimate the 
$V$-band magnitude of GD\,71 gives $V=13.00$, which is close to
$V=13.032\pm0.001$ determined by \citet{landolt92}.


\subsection{Near-IR adaptive optics with NACO}\label{NIR} 

From October 2003 to September 2004, $JHK_sL'$ imaging was obtained
with the NAOS/CONICA (or NACO, for short) equipment, mounted on the
ESO VLT UT4. The $JHK_s$ observations were performed as an extension of
the study previously conducted with the ISAAC instrument
\cite[][ hereafter S03]{sollerman03} 
to study the near-IR properties of the knot.
With the NACO adaptive optics system we were able to obtain
significantly higher spatial resolution. The pixel scale of these
observations were 0.027 arcsec per pixel, and the typical stellar
profile in these images have a FWHM of $\sim0\farcs12$ in $HK_sL'$ 
and $\sim0\farcs25$ in the $J$ band.  
Table~\ref{tab:nacolog} is a log of the NACO observations.
The $J$-band observations, for example, were obtained in sequences of 
$3\times10$ second exposures to make up the total exposure time shown in 
the Table, while the $L'$-band exposures were conducted using a number 
of shorter sequences ($180\times0.175$~seconds).

\begin{figure}
	\resizebox{\hsize}{!}{\includegraphics{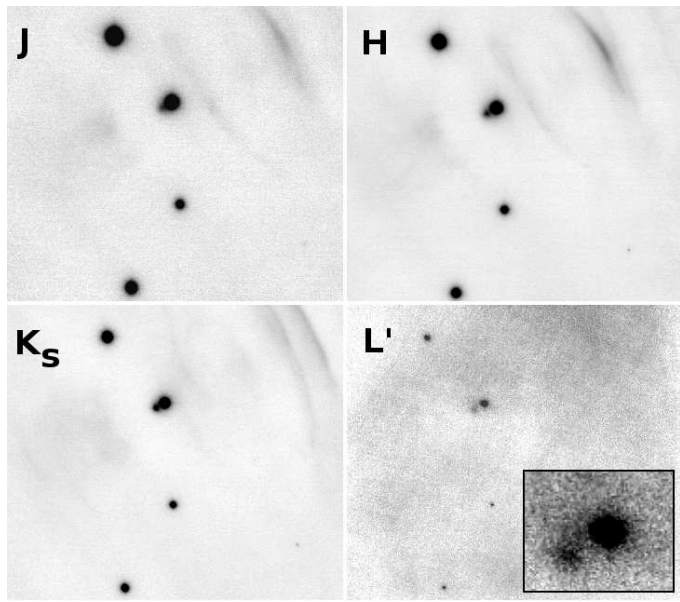}}
\caption{ 
Near-IR NACO images of the Crab Pulsar and its environment. 
North is up and east is to the left. The field of view is
$\sim20\times18$~arcsec. The pulsar is the lower right of the
two bright upper stars. The inlay in the lower right corner 
shows a close-up of the pulsar and the knot in the $L'$ band. 
The field of view for the inset is $\sim2.2\times1.8$ arcseconds, 
and the knot positioned 0.6 arcsec south east of the pulsar is easily resolved.
}
\label{fig:jhkl}
\end{figure}
 
The reductions were performed
using the flat and jitter algorithms from the Eclipse software package
\cite[see][ for a description of Eclipse]{devillard97}. 
We display our near-IR NACO images in Fig.~\ref{fig:jhkl}.

The excellent spatial resolution enable us to reveal the contribution from 
the knot at these wavelengths. To our knowledge, this is the first
ground based $L$-band image presented of the Crab Pulsar. 
Our combined $L'$-band image consists of almost 3 hours of useful 
on-source integration time, and highlights the importance of the 
contribution from the nearby knot at these longer wavelengths
(see the lower right inset of Fig.~\ref{fig:jhkl}).

\begin{figure}
  \resizebox{\hsize}{!}{\includegraphics{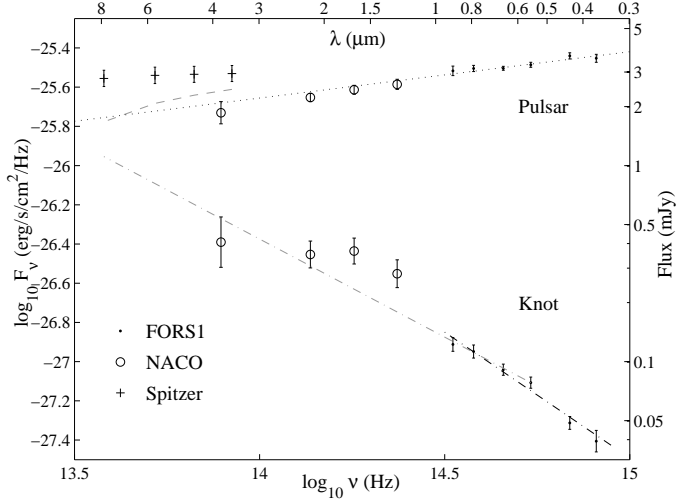}}
  \caption{The measured flux of the Crab Pulsar and its knot. The dotted line shows the least-squares fit to the FORS1 and NACO pulsar data, with $\alpha_\nu=0.27$. The black dash-dotted line shows the fitted line to the optical knot flux, with $\alpha_\nu=-1.3$. The grey dash-dotted line illustrates a hypothetical extrapolation of the knot flux with a spectral index of $\alpha_\nu=-1.0$ into the Spitzer wavelength range, and the grey dashed line shows the resultant knot subtracted Spitzer pulsar flux for this assumption.}
  \label{fig:all}
\end{figure}

Albeit adaptive optics is excellent in obtaining improved spatial resolution, 
it can be difficult to obtain proper absolute flux calibration. 
We have therefore calibrated our NACO 
observations using the known magnitudes of
four stars in the Crab Pulsar field (from S03, their Table~1).
Comparing with the absolute flux calibrated 
NACO images \cite[using standard stars from][]{persson98}, 
this resulted in corrections to the pulsar magnitudes of order 0.06 mag 
in the $JHK$ bands.  
For the $L'$ band, we had to rely on the NACO calibration, 
as obtained from several near-IR standard stars \citep{leggett03}. 
This magnitude estimate thus comes with an additional uncertainty 
of order $\sim0.06$~mags, which is included in Table~\ref{tab:mags} and
Fig.~\ref{fig:all}.

\begin{table}
\centering
\begin{tabular}{c c c}
\hline\hline
Date & Bandpass & Exp. Time (s) \\
\hline 
2003-10-18 & $K_s$ & 1260  \\
2004-01-30 & $L'$  & 1228.5 \\
2004-01-31 & $H$   & 1440  \\
2004-01-31 & $L'$  & 1512  \\
2004-02-12 & $J$   &  990  \\
2004-02-12 & $L'$  & 1197  \\
2004-02-13 & $L'$  & 1512  \\
2004-02-22 & $L'$  & 1512  \\
2004-03-13 & $L'$  & 1512  \\
2004-09-15 & $L'$  & 1512  \\  
\hline
\end{tabular}
\caption{Log of the NAOS/CONICA $JHK_sL'$ observations.}
\label{tab:nacolog}  
\end{table}

\subsection{The Spitzer Space Telescope data}\label{MIR} 

In order to examine the Crab
Pulsar further into the mid-infrared, we have also re-analyzed archival
data\footnote{P.I. Robert D. Gehrz.} 
from the Spitzer Space Telescope. These were acquired with the
IRAC instrumentation on March 6, 2004. 
The IRAC observations were made using 12
second exposures and the high dynamic range mode at four dither
positions, and the data were previously analysed and published by
\citet[ hereafter T06]{temim06}.

Spitzer operates in four channels simultaneously, 
with central wavelengths at approximately 3.6, 4.5, 5.8 and 8.0 $\mu$m. 
For each of the five post-BCD mosaics in each channel, 
we used an aperture radius of 2 pixels (2$\farcs$4), 
with sky annulus radii of 2 to 6 pixels. 
We then applied the appropriate 
aperture corrections listed in the IRAC Data Handbook.

\begin{table}
\caption{
The $UBVRIz$, $JHK_sL'$ and Spitzer IRAC magnitudes of the Crab Pulsar and its knot.}  
\label{tab:mags}            
\centering                      
\begin{tabular}{c c c c}        
\hline\hline
Bandpass & Crab Pulsar + Knot & Crab Pulsar & Knot \\
\hline
$U$ & 16.68$\pm$0.03 & 16.69$\pm$0.03 & 21.57$\pm$0.13 \\
$B$ & 17.22$\pm$0.02 & 17.23$\pm$0.02 & 21.91$\pm$0.08 \\
$V$ & 16.64$\pm$0.02 & 16.66$\pm$0.03 & 20.72$\pm$0.07 \\
$R$ & 16.14$\pm$0.01 & 16.17$\pm$0.02 & 20.02$\pm$0.07 \\
$I$ & 15.61$\pm$0.01 & 15.65$\pm$0.02 & 19.26$\pm$0.08 \\
$z$ & 15.35$\pm$0.04 & 15.39$\pm$0.05 & 18.88$\pm$0.08 \\
$J$   & 14.72$\pm$0.03 & 14.83$\pm$0.03 & 17.24$\pm$0.06 \\
$H$   & 14.13$\pm$0.02 & 14.28$\pm$0.02 & 16.33$\pm$0.04 \\
$K_s$ & 13.64$\pm$0.02 & 13.80$\pm$0.01 & 15.80$\pm$0.03 \\
$L'$  & 12.65$\pm$0.08 & 12.86$\pm$0.07 & 14.51$\pm$0.15 \\
$3.6$ & 12.54$\pm$0.07 & ... & ... \\
$4.5$ & 12.04$\pm$0.07 & ... & ... \\
$5.8$ & 11.55$\pm$0.07 & ... & ... \\
$8.0$ & 10.94$\pm$0.08 & ... & ... \\
\hline
\end{tabular}
\end{table}

\section{Results} \label{results}

We present 
our optical $UBVRIz$, near-IR $JHK_sL'$ 
and the Spitzer Space Telescope magnitudes
of the Crab Pulsar and the knot in
Table~\ref{tab:mags}. Values for the combination of both pulsar
and knot are also presented, for comparison to previous studies
where the two components could typically not be separated.

Compared to previous measurements, 
which have been rather inconclusive 
\citep[see e.g.,][ their Fig.~2]{oconnor05}, 
our $JHK_sL'$ magnitudes agree rather well with the 
time-resolved measurements of \citet{penny82}. 
          
For the Spitzer data, we found considerable differences compared to
T06. This was found to be due to improper aperture corrections in T06, 
and updating their measurements, 
\citet{temim09} now obtain results consistent with ours. 
This changes the slope of the mid-IR measurements significantly.

The spectral energy distribution from the 
measured flux for the Crab Pulsar, including the NACO
and Spitzer data, is shown in Fig.~\ref{fig:all}. 
Here we have corrected the flux 
for interstellar extinction, using the values $E(B-V)=0.52$~mag
and $R_V=3.1$ \citep[][ hereafter S00]{sollerman00}. 
The simplest way to characterize the SED is by a power-law, 
F$_\nu$ = K $\nu^{\alpha_\nu}$, 
where K is a constant and the spectral index is denoted ${\alpha_\nu}$.
If we fit a linear slope to the 
FORS1 and NACO data in Fig.~\ref{fig:all} we get for the knot subtracted pulsar 
a spectral index of $\alpha_\nu=0.27\pm0.03$. 

To estimate the spectral index of the knot we only use the 
optical FORS1 data. 
This is because the knot is known to vary in strength, and the
NACO data are not contemporaneous.
The optical spectral index for the knot obtained in this way is
$\alpha_\nu=-1.3\pm0.1$.


\section{Discussion of SED data}\label{discussion}

The measured spectral index of $\alpha_\nu=0.27$ for the Crab
Pulsar in the optical--near-IR 
wavelength regime agrees reasonably well with previous
studies 
\citep[see e.g. a compilation of results in][ their Table 6]{fordham02}.
The exact value of the spectral index 
depends somewhat on the included wavelength interval, 
and is slightly flatter for the bluer parts of the 
ultraviolet to near-IR regime (S00,~S03).
The value also depends on the applied extinction correction (S00).
The novel thing with these measurements is our attempt to remove the 
contribution from the knot. This will be discussed further below.

The high-resolution NACO data also provide a close view of the pulsar and the knot in the
near-IR (Fig.~\ref{fig:jhkl}). 
These data can be used in several ways.

The spatial extent of the knot can be estimated from our images.
The observed size of $\sim0\farcs25-0\farcs30$ 
corresponds to a unconvolved extension of $\sim0\farcs22-0\farcs27$, which 
is slightly larger
than the estimate by \citet{hester95} in the optical regime.
The center of the knot is located at position angle (PA) 
$\sim$120$^{\circ}$ from the 
pulsar, which agrees well with the knot being aligned
with the rotation axis of the pulsar 
\citep[PA 124$^{\circ}$ $\pm$0.1$^{\circ}$,][]{ngromani04}.
The knot appears slightly elongated along an axis perpendicular to this, with 
PA $\sim$33$^{\circ}$.
As pointed out by \citet{melatos05} it is difficult to measure the exact shape
of the knot in adaptive optics images where the PSF varies across the field.
We do note, however, that the shape of the knot in our NACO 
images is very similar to that presented by 
them \citep[][ their Fig.~7]{melatos05}. 

Our deep $H$-band image was used to establish that any
counter-knot has to be fainter than $\sim2.5\%$ of the knot
luminosity. This is somewhat fainter than the limit 
estimated by \cite{hester95} from HST images,
who used the absence of a counter-knot to argue for an
actual asymmetry in the
polar outflow from the pulsar, rather than simply a Doppler boosting effect.
However, Doppler boosting has more recently been invoked again 
to model the knot in MHD simulations \citep{delzanna06}.

Our measurements of the knot also indicate that it is about twice as bright
compared to the pulsar in the near-IR compared to the ISAAC measurements
of S03. A large variability of this 
structure \citep[S03,][]{melatos05}, as well as other nearby 
structures \citep{hester02} have also been previously reported. 

This variability makes it more difficult to measure the SED of the 
knot itself. The $JHK_sL'$ measurements presented in Table~\ref{tab:mags} 
were obtained at several different epochs, 
which is why we have only connected the contemporaneous 
optical measurements for the knot in Fig.~\ref{fig:all}. 
The SED we measure from the optical data is
$\alpha_\nu=-1.3\pm0.1$. 
This is a bit redder than the earlier 
measurement of $\alpha_\nu=-0.8$ by S03. 

We note, however, that while the measurements by S03 made use of HST images, 
our
FORS images - even though they have the highest quality achievable
from the ground - require PSF subtractions that may introduce
uncertainties in the measurements. The formal fit error of 0.1 mag may thus underestimate the uncertainties.
Having said this, we still
speculate if the suggested difference in spectral slope for the knot
between these two measurements may be related to the variability of
the knot brightness. A multi-band monitoring campaign with the updated 
HST would be able to investigate such an energy injection emission
mechanism for the knot 
(see Sect.~\ref{sect:outlook} for a discussion on future observations).

The observations of S03 did show that
the emission of the knot rises strongly into the
infrared, and that was part of the motivation for this renewed study. 
While the knot contributes only $\sim1$\% in the $U$ band, this knot/pulsar 
ratio increases to $\sim4$\% in the $z$ band. The NACO study clearly shows
that the knot is even more important in the $L'$-band regime,
where the knot-to-pulsar ratio amounts to 22$\pm$8\% at this epoch.

Our measurements on the Spitzer data show that the pulsar emission
continues with a flat energy distribution all the way out to the
longest wavelengths spanned (Fig.~\ref{fig:all}).
The spatial resolution of these data do not allow us to separate 
the contribution of the pulsar and the knot.
The strong contribution of the knot in the $L'$ band suggests
that the knot emission could continue well into the 3-8 $\mu$m 
Spitzer range. This could explain part of the apparent 
offset for the Spitzer data in Fig.~\ref{fig:all}. 
We discuss this further below.

The approximate agreement of 
the measured spectral index of the pulsar 
in the optical -- near-IR with $\alpha_\nu\sim1/3$ expected for 
optically thin synchrotron radiation, is a valuable clue for 
understanding the emission mechanism of pulsars. 
However, for the longest wavelengths of measured non-thermal 
incoherent emission, much of the discussion has traditionally focused on the 
indication of a possible self-absorption roll-over in the
infrared \citep[see references in][]{oconnor05}. 
Our new time-averaged pulsar measurements 
do not indicate a simple synchrotron self-absorption cut-off 
(see further discussion in Sect.~\ref{sect:SSA}).

\section{The Secular Decrease in Luminosity}\label{secular} 

Having investigated the SED, we now turn to the 
possible time evolution of the optical luminosity. 

A decrease in the pulsar luminosity by $\sim$5 milli-magnitudes (mmag) 
per year was predicted by \citet{pacini71}. 
This secular decrease in luminosity 
has since been repeatedly claimed, invoked or discussed
in the literature 
\cite[e.g.,][]{kristian78, penny82, pacinisalvati83, middleditch87, 
shearer02, shearer08}. 

The most recent measurement of this decrease in luminosity is from
\citet{nasuti96}. They estimated a decrease of 
$8\pm4$~mmag~yr$^{-1}$ in the $V$ band, but
also caution that more observations are needed. 

Here we present measurements based on data from the Jacobus
Kapteyn Telescope (JKT), 
the Nordic Optical Telescope (NOT) and the VLT.
We concentrate on images obtained in the $V$-band over the
past 20 years. In this way we are able to increase the number of observations
as compared to \citet{nasuti96} by a factor of 4,  
and decrease the individual error bars by a factor of 5. 
More importantly, this
investigation 
attempts to minimize the systematic uncertainties. Where \citet{nasuti96} used
a mix of photo-electric time resolved photometry, CCD imaging and photometry
derived from a spectrum, we evoke only modern
CCD imaging at decent seeing and perform relative photometry against several field stars.

\subsection{Predicted secular decrease}\label{predictsecular} 

The predicted decrease in luminosity is based on the simple assumption that
the luminosity (L) of the pulsar depends on the pulsar period (P) as
\[ L\propto\ P^{-n} \] 
\citet{pacini71} argues that the exponent in the synchrotron luminosity
should be $n\sim10$ for the Crab Pulsar, and 
estimated this effect to a secular decrease of 0.005 mag per year. 
The slightly updated analysis in 
\citet{pacinisalvati83} argues that this should be observable in 
the optical pass-bands if the emission is incoherent synchrotron radiation.
Using their Equation 10 for optically thin synchrotron emission 
($\alpha_\nu\sim1/3$), we estimate, using modern values for the pulsar period 
and period derivative, a predicted 
secular decrease of 3.8 mmag per year.

\subsection{Observations of the $V$-band luminosity}\label{observesecular} 

We have gathered archival data from several telescopes.
We have used JKT data for 11 epochs between 1988 and 2000, where 
some of these data were previously used to investigate the 
wisp motions in the Crab nebula 
\citep{tanvir97}. We have also used
data from
the NOT from 1998, 2000 and 2008, 
the abovementioned FORS1 data from December 2003
as well as additional FORS1 data from November 2003, and FORS2 data
from 2005. All these observations were performed in the $V$-band filter.
A log of the observations is shown in Table~\ref{tab:jktlog}. 
These data include only images with seeing better than 1.5 arcsec, in order to
minimize errors in background subtraction.

All data were reduced in a standard way. 
The relative photometry between the Crab Pulsar and 8 stars in the field 
was performed with PSF photometry, in order to minimize 
the contribution from the nebula, the wisps and the knot. 

To get the zero-point for each exposure we fixed the magnitude of 
the field star positioned 6$\farcs$4 north, 20$\farcs$6 east of the pulsar 
(Star 1 in Table~\ref{tab:trendmags} and Fig.~\ref{fig:trendstars}). 
Of the 8 well exposed and relatively 
isolated field stars that we have used, 
this is the one with $V$ magnitude most 
similar to that of the Crab Pulsar, it is close to the pulsar position 
and was also visible in all images used. 
The root-mean-square (rms) of the difference in relative photometry for the field stars 
was then used for estimating the uncertainty in each epoch. 

We find that the local standard stars have an rms scatter of 0.02 mag 
over the 17 used epochs. 
This is thus the precision in photometry we 
can reach in this study. This is larger than the typical photon noise 
errors in these frames,  and is probably mainly limited by flat fielding 
errors, but also the neglect of color terms and on the nebular background 
combined with varying seeing conditions.

\subsection{Secular Results}\label{resultssecular} 

\begin{figure}
  \resizebox{\hsize}{!}{\includegraphics{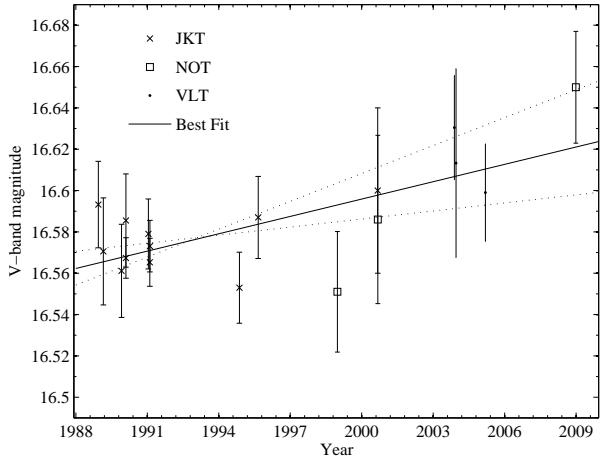}}
  \caption{Measured $V$-band magnitude of the Crab Pulsar from our investigation of the secular decrease. 
Crosses mark JKT data, squares NOT data and dots VLT data. The solid line shows the best fit of $2.9$ mmag/year, with the estimated uncertainties marked by the dotted lines.
}
  \label{fig:jktpsf}
\end{figure}

The results from our study of the secular decrease in luminosity are
shown in Fig.~\ref{fig:jktpsf}. The best (linear least-squares) fit to our
data yields an increase of $2.9\pm0.8$ mmag/year. This
purely statistical error likely underestimates the uncertainties. 
We therefore use our 7 independent field stars for comparison, and fit linear
slopes also to these. While they rightly show
zero evolution on average, 
the rms scatter on the slopes is 1.4 mmag/year. 
We therefore add this to the uncertainty above, arriving at a
final estimate of $2.9\pm1.6$ mmag/year. 
This is thus a $2\sigma$ detection of a secular 
decrease.

\section{Summary, Discussion and Outlook}\label{conclusions}

We have measured the optical to infrared SED of the Crab Pulsar and its 
nearby knot. For the pulsar itself, we find a relatively flat spectrum over the 
entire wavelength range. 
The spectral index in the optical--near-IR is 
$\alpha_\nu = 0.27\pm0.03$. 
The nearby knot has a much redder spectrum. 
We have found that the intensity of the knot can vary by a factor 2. 
In our $L'-$band measurements, the knot contributes almost 20\% of 
the knot+pulsar emission.

In this investigation we have scrutinized two often invoked concepts 
for the Crab Pulsar, the infrared synchrotron self-absorption (SSA) roll-over, 
and the secular decrease in luminosity. 

\subsection{Synchrotron Self-absorption}\label{sect:SSA}

The spectral index of the optical to infrared regime has been 
discussed in many papers 
\citep[e.g., S00; S03;][]{middleditch83,penny82,temim06}.  
\citet{oconnor05} review the evidence for a SSA roll-over, 
which is the very basis for their discussion. The notion of an IR SSA is even
included in the text books on pulsars \cite[][ chapter 19.5]{lynegrahamsmith}.
In simple standard theory, such a SSA break would give rise to the longer 
wavelengths falling off with  $\alpha_\nu\sim2-2.5$.

Looking at our time-averaged 
data, we see in fact no evidence that the slope is 
steepening away from the $\alpha_\nu=1/3$ expected from simple optically 
thin synchrotron radiation. 
In particular, the Spitzer data reveal a flat spectrum
all the way out to 8 microns, which does not confirm earlier claims 
\cite[e.g.,][]{middleditch83}.  
Our main conclusion is therefore that there is simply little support for the 
SSA interpretation in the IR.

We end this section by noting that
the nearby knot displays a much redder
spectrum, with a spectral index 
reminiscent of that 
expected from Fermi-acceleration and cooling behind a shock.

In Fig.~\ref{fig:all} we show that 
if such a $\alpha_{\nu}^{\rm knot} = -1.0$ extrapolates into
the Spitzer region, the knot can indeed contribute substantially in this
regime. This could perhaps make the pulsar SED consistent with a 
$\alpha_\nu\sim1/3$ slope 
all the way from the UV regime. If the knot is even stronger,
the emission detected by Spitzer
would in fact mostly be due to the knot, 
while the pulsar itself must be suppressed by the SSA. 
Alternatively, the 
cooling break for the knot occurs before the mid-IR.
These speculations can in fact be tested by future observations (Sect.~\ref{sect:outlook}).

\subsection{Secular Decrease}
We have also investigated the secular decrease of the optical 
emission of the Crab Pulsar.
We note that this effect has been extensively invoked in the literature, but that the observational evidence so far has been meager. In fact, we claim that our investigation is by far the most reliable one to date, but still only  
indicate a secular decrease at the $2\sigma$ level. 
Our measurement agrees well with an updated Pacini's Law, 
predicting a decrease of $\sim4$ mmag/year.

We believe that 
better estimates of the secular decrease can be obtained (see below).
This can not only confirm (or reject) this decrease, but can in detail probe
the luminosity evolution of the pulsar and thus the emission mechanism. This
is clearly a more direct test on the pulsar evolution than attempts to monitor
the synchrotron emission from the entire nebula \citep[e.g.,][]{aller85,reynoldschevalier84,smith03}.
For example, in the original L = Const.~$\times$~P$^{-\rm{n}}$ 
formulation of \cite{pacini71}, our current measurement indicates
$n=6.8\pm3.8$. 
While current pulsar emission theory is not yet 
very powerful in predicting the exact numbers 
\cite[e.g.,][]{smith81,oconnor05}, 
it is clear that better measurements will be able to
discriminate between the $n=-10$ scenario favoring emission 
from the proximity of the speed-of-light cylinder, or a 
significantly shallower 
index  $n=-4$ if the optical flux scales with the total loss of 
rotational energy from the neutron star.

\subsection{Outlook}\label{sect:outlook}

New observations will be able to resolve several of the above-mentioned issues. Imaging from space (e.g., with the HST) can be 
used to investigate the optical SED (and any time evolution) of the knot
and of the pulsar. 
In particular, the JWST with sub-arc-second resolution
over a large IR wavelength range will be ideal to resolve the knot
at the longest wavelengths, and will determine to 
what extent the knot contributes.

Much of the discrepancies found in published values of the pulsar SED
comes from the efforts of patching together broad-band magnitudes
obtained under varying conditions and with very different
instrumentation. 
The next step to get the pulsar+knot SED would be to use the
X-shooter instrument at the VLT to directly probe the spectrum all the
way through the optical-NIR.

For the secular decrease it is also evident 
that more high quality observations 
will be able to more clearly investigate this effect. 
Decreasing the errors by a factor 3 would enable interesting 
comparison to theory, and this is achievable with a uniform dataset.
The best
possible dataset would of course have been precision measurements with
the HST, where a lot of attention has already been directed
to studies of the time evolution of the pulsar environment
\cite[e.g.,][]{hester95,hester08}. 
Unfortunately, the pulsar itself 
has been categorically saturated in these investigations
\cite[see][]{kaplan08}, and the existing HST observations are therefore
not suitable for investigating the pulsar SED or secular decrease.
Other datasets obtained 
with a uniform telescope/instrument set-up, 
as e.g., the images
used for investigating the expansion of the nebula over time
\citep[e.g.,][]{nugent98,trimble68, wyckoffmurray77} could also be of interest. We note that there must exist significantly more data
which could be useful for this type of analysis. 
We therefore encourage the astronomical community to investigate or share their
Crab Pulsar images.

\begin{acknowledgements}
This paper is based on results from the master thesis of Andreas Sandberg.
Jesper Sollerman is a Royal Swedish Academy of Sciences Research Fellow 
supported by a 
grant from the Knut and Alice Wallenberg Foundation. 
The Oskar Klein Centre is funded by the Swedish Research Council (VR), and 
A. Sandberg acknowledge further support from VR.
We thank the anonymous referee for a concise report.
We thank Tea Temim for a fruitful conversation on the Spitzer data reductions, 
Yura Shibanov for pointing us
in that direction and N. \& P. Lundqvist for early discussions on this project.
Claes-Ingvar Bj\" ornsson is thanked for many discussions on the theoretical aspects of this work, and we thank 
M. Persson for the Albanova Telescope images
and W. Nowotny for the NOT images from 2000.  
Some of the data presented here have been taken using ALFOSC, which is owned by the Instituto de Astrofisica de Andalucia (IAA) and operated at the Nordic Optical Telescope under agreement between IAA and the NBI in Copenhagen
This paper also makes use of data obtained from the Isaac Newton Group Archive which is maintained as part of the CASU Astronomical Data Centre at the Institute of Astronomy, Cambridge. 
This work is based in part on observations made with the Spitzer Space Telescope, which is operated by the Jet Propulsion Laboratory, California Institute of Technology under a contract with NASA.
\end{acknowledgements}

{}

\bibliographystyle{aa} 

\onltab{3}{
\begin{table*}
\centering                       
\begin{tabular}{c c c c c c}
\hline\hline
      Date     &    Telescope    &   Instrument/Detector & Exposure Time (s)  & Seeing &  Stars Not Included  \\
      \hline
  1988-12-11   &         JKT      &          GEC4       &     7 $\times$ 1000 & 1$\farcs$4 &            4,5\\
  1989-02-26   &         JKT      &          GEC3       &     2 $\times$ 1000 & 1$\farcs$4 &              5\\
  1989-12-02   &         JKT      &          RCA2       &     3 $\times$ 1000 & 1$\farcs$3 & ...\\
  1990-02-07   &         JKT      &          GEC3       &     2 $\times$ 1000 & 1$\farcs$1 & ...\\
  1990-02-08   &         JKT      &          GEC3       &     5 $\times$ 1000 & 1$\farcs$1 & ...\\
  1991-01-17   &         JKT      &          GEC3       &     3 $\times$ 1000 & 1$\farcs$2 & ...\\
  1991-02-09   &         JKT      &          GEC3       &     4 $\times$ 1000 & 1$\farcs$4 &              2 \\
  1991-02-10   &         JKT      &          GEC3       &     3 $\times$ 1000 & 1$\farcs$2 &              2 \\
  1994-11-14   &         JKT      &          TEK4       &     1 $\times$ 1200 & 1$\farcs$3 &   5(S),6(S),7(S)\\
  1995-08-28   &         JKT      &          TEK4       &     1 $\times$ 300  & 1$\farcs$4 & ...\\
  1998-12-25   &        NOT       &         LORAL       &     1 $\times$ 30   & 1$\farcs$4 & ...\\
  2000-09-03   &         JKT      &          SITe2      &     1 $\times$ 250  & 1$\farcs$1 & ...\\
  2000-09-04   &        NOT       &         LORAL       &     1 $\times$ 300  & 0$\farcs$9 & ...\\
  2003-11-23   &         VLT      &         FORS1       &     2 $\times$ 5    & 0$\farcs$7 & ...\\
  2003-12-18   &         VLT      &         FORS1       &     3 $\times$ 20   & 0$\farcs$6 & ...\\
  2005-03-12   &         VLT      &         FORS2       &     1 $\times$ 0.987  & 1$\farcs$3  &       5,9 \\
  2008-12-19   &        NOT       &         EEV        &     3 $\times$ 180 & 0$\farcs$9 & ...\\
\hline 
\end{tabular}
\caption{$V$-band data used for examining the secular decrease in luminosity of the Crab Pulsar.
The final column lists stars that are saturated (S) or outside the field of view in the particular image, and the numbers refer to Table~\ref{tab:trendmags} and Fig.~\ref{fig:trendstars}. 
} 
\label{tab:jktlog}        
\end{table*}
}

\onlfig{4}{
\begin{figure*}
  \resizebox{\hsize}{!}{\includegraphics{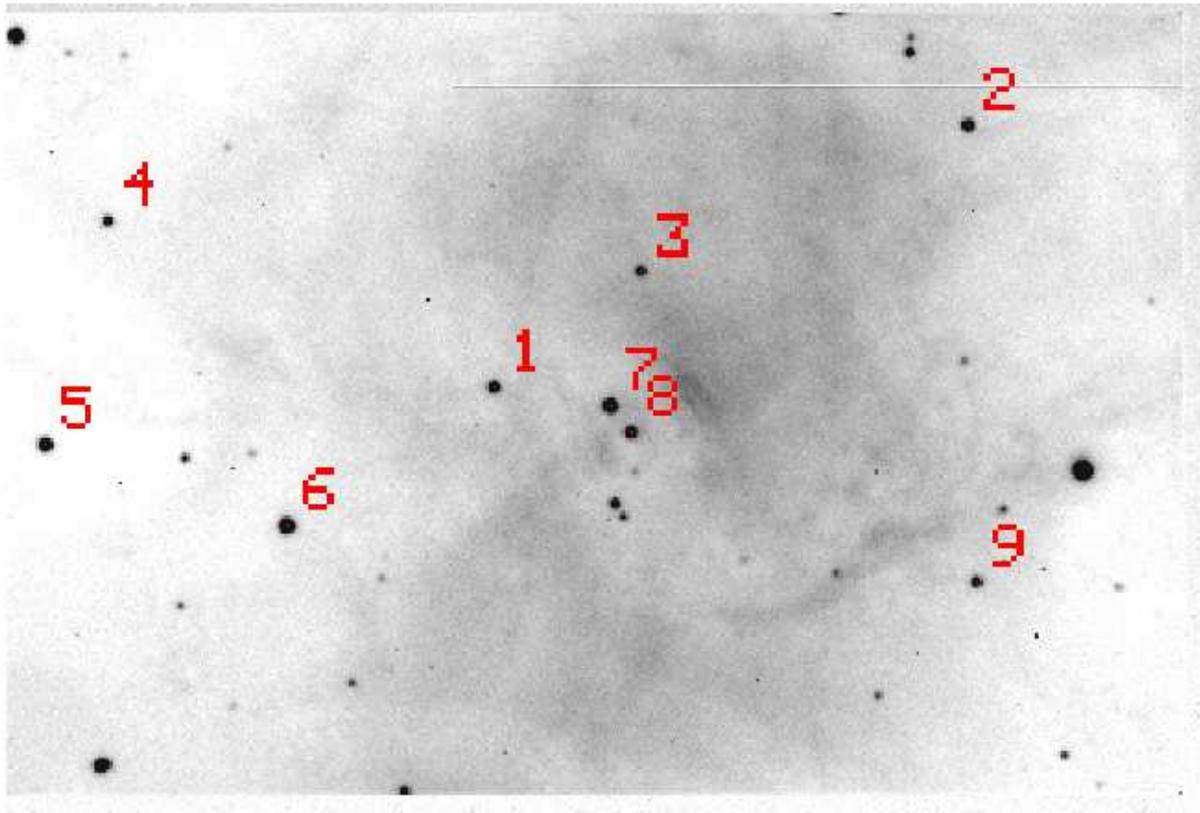}}
\caption{The stars that were chosen for relative photometry in the investigation of the secular decrease in luminosity. The numbering is the same as in Table~\ref{tab:trendmags} where the $UBVRIz$ magnitudes for these stars are given. The numbering is also used in Table~\ref{tab:jktlog}. The pulsar is number eight. This $V$-band image was acquired with the JKT on February 7 1990. North is up and east is to the left. The field of view is approximately 3.2 by 2.1 arcminutes.}
\label{fig:trendstars}
\end{figure*}
}

\onltab{4}{
\begin{table*}
\centering                       
\begin{tabular}{c c c c c c c}
\hline\hline
Star No. & U & B & V & R & I & z \\
\hline
1 & 17.80 & 17.49 & 16.51 & 15.92 & 15.36 & 15.13 \\
2 & 18.43 & 17.74 & 16.36 & 15.59 & 14.87 & 14.53 \\
3 & 18.42 & 18.29 & 17.40 & 16.82 & 16.26 & 16.02 \\
4 & 18.74 & 18.32 & 17.16 & 16.48 & 15.86 & 15.59 \\
5 & 18.45 & 17.23 & 15.64 & 14.80 & 14.07 & 13.71 \\
6 & 16.56 & 16.29 & 15.37 & 14.78 & 14.22 & 13.98 \\
7 & 16.80 & 16.64 & 15.79 & 15.25 & 14.74 & 14.52 \\
9 & 18.13 & 17.93 & 17.07 & 16.51 & 15.97 & 15.75 \\
\hline
\end{tabular}
\caption{$UBVRIz$ magnitudes of the eight field stars examined in the secular decrease of luminosity of the pulsar. The typical errors in $VRI$ are 0.02 mags, while $B$, $z$ and $U$ give uncertainties of 0.03, 0.04 and 0.05 respectively. These magnitudes were estimated with the FORS1 calibrations described in Section~\ref{optical}. The numbering corresponds to Fig.~\ref{fig:trendstars} and Table~\ref{tab:jktlog}.}
\label{tab:trendmags}
\end{table*}
}

\end{document}